# FIRST-PRINCIPLES STUDY ON CONTROLLING ENERGY GAP OF GRAPHENE USING HYBRID ARMCHAIR-ZIGZAG NANOSTRUCTURES

NGUYEN TIEN CUONG

*Faculty of Physics, VNU University of Science*
*334 Nguyen Trai, Thanh Xuan, Hanoi, Vietnam*
[ntcuong@vnu.edu.vn](mailto:ntcuong@vnu.edu.vn)

The electronic and transport properties of hybrid armchair-zigzag nanostructures including U-shaped graphene nanoribbons (GRNs) and patterned nanopores structured graphene were studied using combination of density functional theory and non-equilibrium Green's function method. The density of state, electron transmission spectra, and molecular orbitals were analyzed. The obtained results show that GNRs junctions tend to open an energy gap when U-shaped structures were formed due to the formation of quasi-bound states at zigzag edges. The size of U-shaped structures has enormous influences on the electron transport of the system. We also considered the effect of corner form of the U-shaped GNRs junctions on energy gap opening. It was found that as some carbon atoms are add to the inner corner, the energy gap in U-shaped GNRs significantly changed. For patterned nanopores structured graphene, the calculated results show that patterned nanopores enormous influence on electronic and the transport properties though the GNRs junctions, depending on the shape, size, and the number of nanopores. The study suggests that designed tailored graphene systems based on hybrid armchair-zigzag nanostructures can be used to control the energy gap of graphene.

## 1. Introduction

The discovery of graphene and its successful fabrication in 2004 [1-4] has opened up a new field of fundamental physics for new electronic devices. Based on all its advantages, graphene is a potential candidate that could replace silicon based materials for nano-scale devices. However, there are a number of problems with grahene. Unlike silicon, pristine graphene does not have an energy gap which is essential for semiconductor device operation.

Carbon nanotubes (CNTs) are graphene rolled up into a tube. CNTs can be either metallic or semiconducting depending on their charities. Because the symmetric and unique electronic structure of graphene and a structure of rolled nanotube strongly affects its electronic properties. It implies that band gap of CNTs can be modified by controlling the charities of them. In another way, band gap of graphene can be created by cutting graphene into graphene nanoribbons (GNRs). There are two kinds of GNRs named Armchair GNRs (AGNRs) and Zigzag GNRs (ZGNRs). ZGNRs are metallic for all widths of ribbons while AGNRs are either metallic or semiconducting depending on their widths [5-7]. The hybrid AGNR-ZGNR junctions possess electronic characteristics of both AGNR and ZGNR. However, the electronic properties of the single kinds of GNR junctions are much different from those of the hybrid junctions. Therefore, the hybrid junctions have been nanostructures of strong interest. Recently, various hybrid AGNR-ZGNR junctions, such as L-shaped, Z-shaped, T-shaped and U-shaped have been studied [8-14]. Furthermore, electronic structures of twisted GNRs [15], and strained GNRs [16, 17] have been also examined. It is show that electronic and transport properties of these GNRs junctions have been significantly modified in comparison with those of pristine structures. Furthermore, both experiment and calculation demonstrated that bilayer graphene gets a band gap by applying a strong electric field in the direction perpendicular to the bilayer graphene plane [18-20]. All the results suggested that geometrical structures of graphene could play key roles in controlling band gap of these systems.

Motivated by successful fabrications U-shaped graphene channel transistors [21], the electronic properties and quantum transport of U-shaped GRNs have been calculated and reported in our previous work [14]. The calculated results shown that the GNRs tend to open a band gap when U-shapes are formed due to the formation of quasi-bound states at zigzag edges. Geometrically, U-shaped GNRs is a kind of the hybrid AGNR-ZGNR nanostructures. In this work, first-principles calculations have been performed to investigate the controlling energy gap of graphene using hybrid amrchair-zigzag nanostructures.

## 2. Model and Computational Details

The electronic and transport properties are calculated using the combination of the density functional theory (DFT) and the non-equilibrium Green's function (NEGF) methods. In particular, the electronic properties such as density of state (DOS), molecular orbitals (MOs), adsorption energy were calculated based on the DFT using both Materials Studio/Dmol3 and OpenMX packages [22, 23]. The transport properties such as transmission spectrum, current-voltage characteristics were calculated by using NEGF method which is implemented in the OpenMX package [23].

For OpenMX calculations, a system consisting of a central region connected to left and right leads with infinite size, as shown in Fig. 1, is treated by the NEGF method. By considering the two dimensional periodicity in the bc-plane, the system can be cast into a one-dimensional problem. The electronic transport is assumed to occur along the a-axis. All calculations were carried out with in the LDA exchange-correlation functional. The SCF energy convergence criterion is set to $10^{-5}$ eV. In the treatment of pseudo-potentials, the pseudo atomic orbital basis sets of C4.5-s2p1 for carbon and H4.5-s2 for hydrogen were used, where in the abbreviation of basis functions such as C4.5-s2p1, C stands for the atomic symbol, 4.5 the cutoff radius (Bohr) in the generation by the confinement scheme, and s2p1means the employment of two primitive orbitals for s and one primitive orbital for p [23].

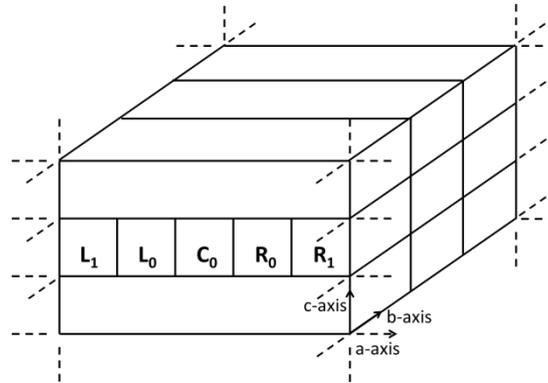

Fig. 1. Configuration of the system treated by NEGF method.

For Dmol3 calculations, the MOs are calculated at the gamma point using the local density approximation (LDA) for the exchange-correlation functional. All-electron calculations were performed with a finite basis set cutoff of 3.5 Å, and self-consistent field (SCF) tolerance of $10^{-5}$ Ha.

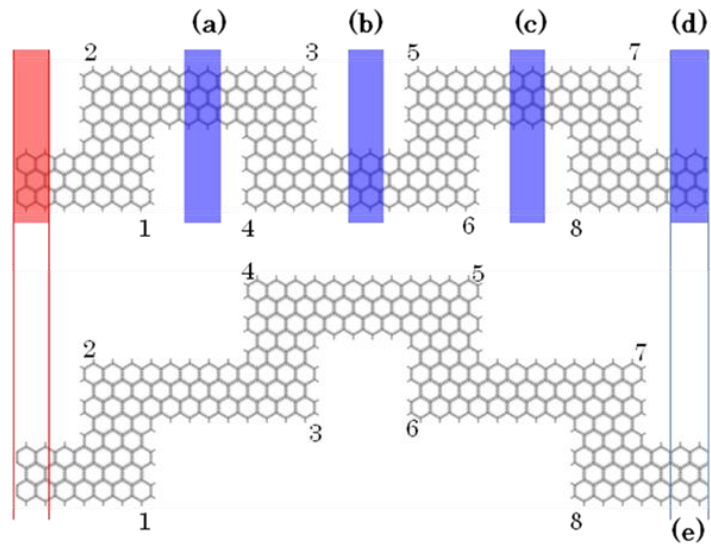

Fig. 2. Configuration of right-angles shaped GNRs structures with left lead (red) and right leads (blue):
(a) 2 right-angles structure, (b) 4 right-angles structure, (c) 6 right-angles structure,
and (d)-(e) 8 right-angles structures.

The right-angles shaped GNRs junctions, as shown in figure 5.9, were used for first-principle calculations. In the model, semi-infinite ZGNRs leads play as left electrode (red region) and right electrode (blue region). The edges of GNRs are terminated by hydrogen.

Fig. 3 shows the configuration of U-shaped ZGNRs junctions with inner corner. The edge structures of inner corner can be zigzag or armchair depending on the angle between the inner edge and vertical edge. For example, in the U-shaped ZGNR case, the angle is 30º corresponding to zigzag edge corner. The angle is 60º corresponding to armchair edge corner.

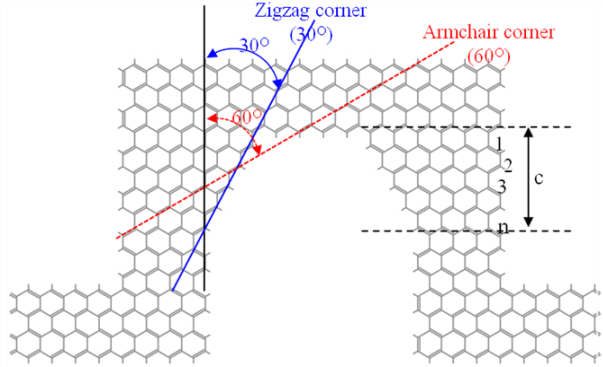

Fig. 3. Configuration of U-shaped GNRs junctions with inner corner.

## 3.   Results and Discussion

### *3.1. The effect of right-angle numbers on energy gap*

A right-angle GNR junction, including both AGNR and ZGNR edges, is a simplest hybrid AGNR-ZGNR junction. The numbers of right-angles effects on their electronic transport properties are examined.

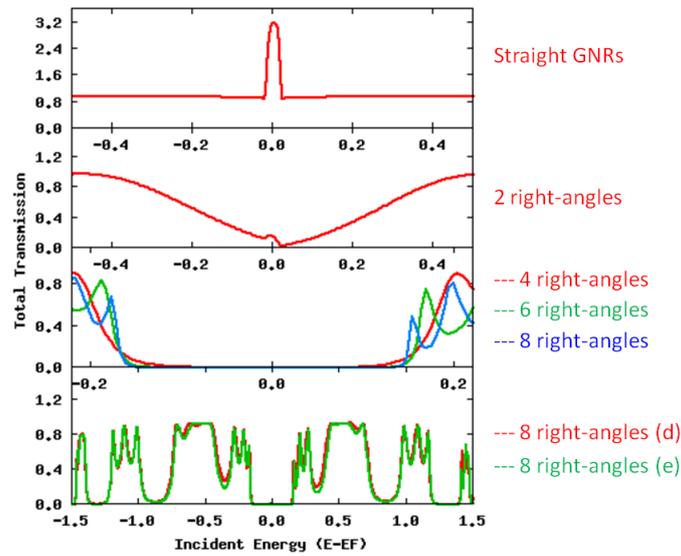

Fig. 4. Total transmission spectrum of of right-angles shaped GNRs, which is shown in the Fig. 2.

Fig. 4 shows the transmission spectra of various right-angles shaped GNR junctions. It is see that right-angles shaped GNR junctions tend to open band gap when the numbers of right-angles are greater or equal to 4 (corresponding to U-shaped GNR junction). The band gap increases insignificant as the numbers of right-angles are greater than 4. It implies that U-shaped GNRs is a potential structure for opening band gap. The obtained results are in good agreement with our previous results [14].

### *3.2. The effect of inner corner size on energy gap*

The effect of inner corner form on electron transport of U-shaped GNRs junctions in both armchair and zigzag cases are investigated. Fig. 5 and Fig. 6 show the transmission spectra of U-shaped ZGNRs and AGRNs junctions with various inner corner forms, respectively.

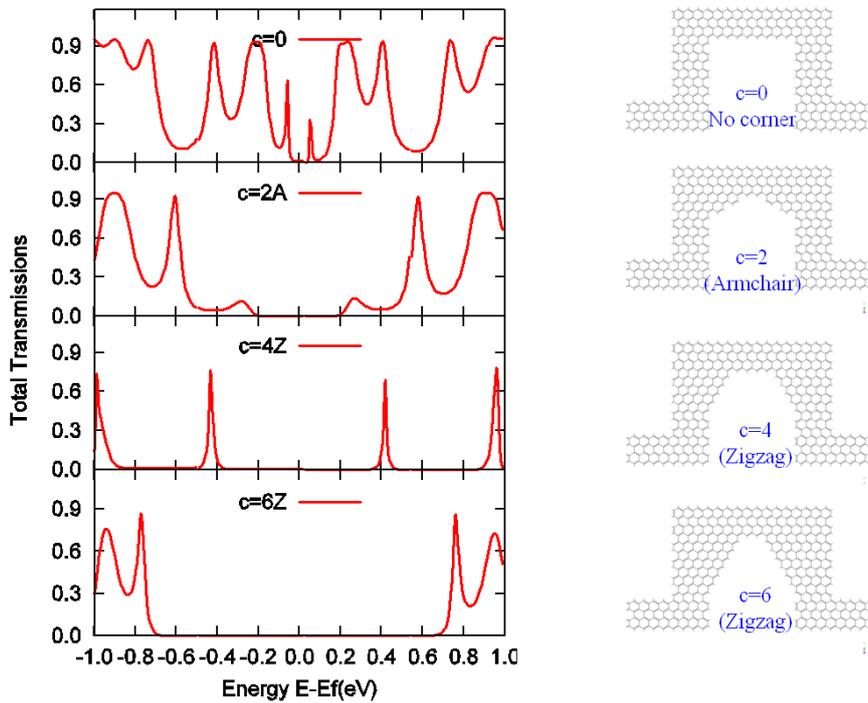

Fig. 5. Transmission spectra of U-shaped ZGNRs junctions with various inner corners

One can see that as carbon atoms are add to the inner corner, the energy gap in U-shaped GNRs increases in both zigzag and armchair cases. The increasing energy gap effects of zigzag edge corners are significantly larger than those of armchair edge corners due to the localized states at zigzag edges of corners. The results suggest that the inner corners of U-shaped GNRs can be used to control the band gap energies.

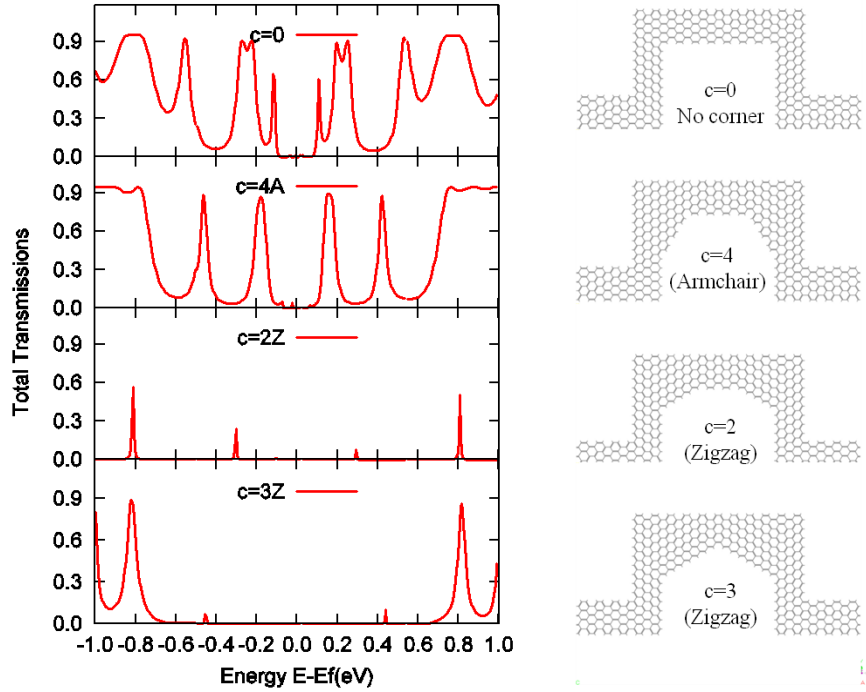

Fig. 6. Transmission spectra of U-shaped AGNRs junctions with various inner corners

### *3.3. The effect of patterned nanopores structurer on energy gap*

The electronic and transport properties of a junction with configuration of one-nanopore structured graphene system have been studied. Fig. 7 shows density of state, HOMO, LUMO, and corresponding transmission spectrum of AGNR with one circular nanopore, square nanopore, and triangular nanopore, respectively. In case of circular nanopore, one can see that the HOMO-LUMO states remain delocalized. However, sharp peaks at the HOMO and LUMO energies of pristine AGNR cannot be seen in DOS, and the transmission rate is largely reduced near the both HOMO and LUMO edges. In case of square nanopore, there are two additional DOS peaks around Fermi level due to localized states at zigziag edges of square nanopore. They don't reflect in transmission peaks, however. Because, there is no corresponding states at the left and right leads. For triangular nanopore, both HOMO and LUMO states become localized due to formation of quasi-bound states at all three zigzag edges of triangle pore. The localized HOMO-LUMO states are dominant around Fermi level. Similar to DOS of circular pore case, sharp peaks at the HOMO and LUMO energies cannot be seen in DOS. Moreover, the sharp peaks at lower HOMO states (HOMO-1, HOMO-2,…) and upper LUMO states (LUMO+1, LUMO+2,…) become wider that those of pristine AGNR. Therefore, the transmission significantly reduces without anti-resonant peaks.

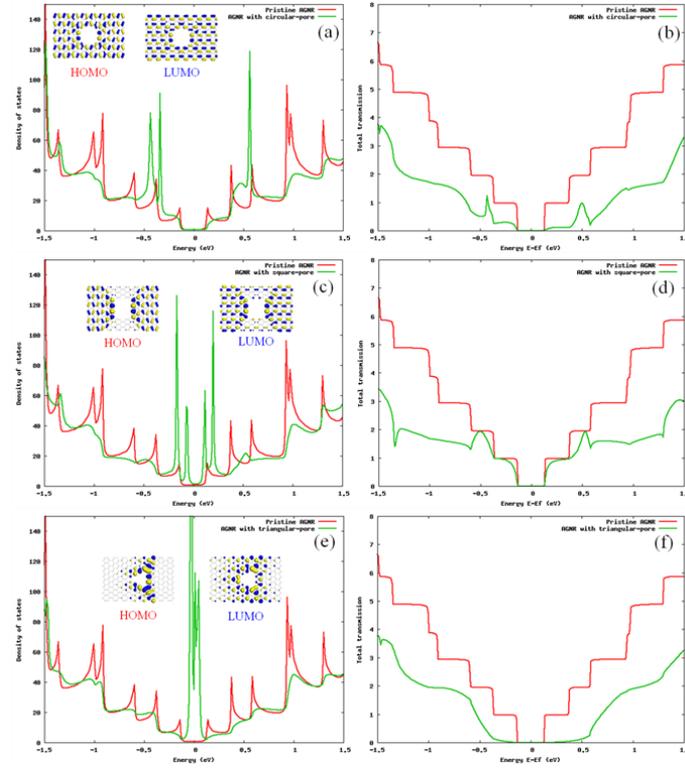

Fig. 7. (a,c,e) Density of state and HOMO-LUMO, (b,d,f) corresponding transmission spectrum of AGNR with one circular nanopore, square nanopore, and triangular nanopore, respectively.

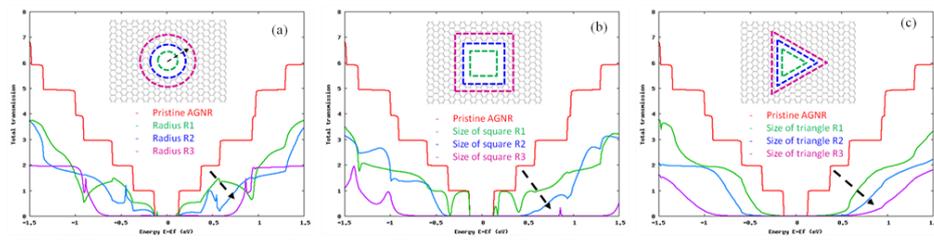

Fig. 8. Transmission spectra and corresponding configuration of one-nanopore structured graphene system with size of the nanopore increase: (a) circular pore, (b) square pore, and (c) triangular pore

Fig.8. show transmission spectra and corresponding configuration of one-nanopore structured graphene system with size of the nanopore increase for (a) circular pore, (b) square pore, and (c) triangular pore. We can see that with the increasing of nanopore size, energy gap of the graphene junction increase in all three cases. Among them, the band gap in case of triangular nano increase without anti-resonant in transmission. Consequently, the obtained results shown that patterned nanopores enormous influence on the electronic and transport properties, depending on the shape, and size of nanopore.

## 4. Conclusions

The electronic and transport properties of hybrid Armchair-Zigzag nano structured systems including U-shaped GNRs based systems and patterned nanopores structured systems were studied using combination of DFT and NEGF methods. We found that the GNRs junctions tend to open a band gap when U-shaped structures were formed in both armchair and zigzag cases due to the formation of quasi-bound states at zigzag edges. The size of U-shaped structures has enormous influences on the electron transport of the system. We also considered the effect of corner form on electron transport of U-shaped GNRs junctions. It is found that as some carbon atoms are add to the inner corner, the energy gap in U-shaped GNRs increased in both Zigzag and Armchair cases. The increasing energy gap effect of zigzag edge corner is significantly larger than that of that of armchair edge corner. For patterned nanopore structured graphnene systems, the calculated results show that patterned nanopore enormous influence on the electronic and transport properties though the GNRs junctions, depending on the shape, and size of nanopores. The obtained results suggest that the designed tailored shape graphene systems based on hybrid AGNR-ZGNR nanostructures can be used for controlling the energy gap of graphene.

**Acknowledgments**

The author thanks the Vietnam National University (VNU) Hanoi for funding this work within project No. QG.17-12. The computations presented in this study was performed at the Faculty of Physics of VNU University of Science, Vietnam, and the Information Center of Japan Advanced Institute of Science and Technology (JAIST), Japan. The author acknowledges Prof. Hiroshi Mizuta and Prof. Dam Hieu Chi, of JAIST, for their useful discussions and suggestions about the research.